\documentclass[%
preprint,
 amsmath,amssymb,
 aps,
prb,
floatfix,
]{revtex4-2}

\usepackage{graphicx}
\usepackage{natbib}
\usepackage{dcolumn}
\usepackage{bm}
\usepackage{hyperref}

\begin{document}

\title{Phonon scattering and vibrational localization in 2D embedded nanoparticle composites}

\author{Ongira Chowdhury}
\affiliation{%
Department of Mechanical Engineering, University of Delaware, Newark, DE, 19716 USA
}%
\author{Joseph P. Feser}%
 \email{jpfeser@udel.edu}
\affiliation{%
Department of Mechanical Engineering, University of Delaware, Newark, DE, 19716 USA
}%

\date{\today}

\begin{abstract}

In this work, a Landauer approach enabled by the Frequency Domain Perfectly Matched Layer Method (FDPML) is used to study phonon transport in a series of large 2D domains with randomly embedded nanoparticles over a wide range of nanoparticle loadings and wavelengths.  The effect of nanoparticle packing density on the mean free path and localization length is characterized. We observe that in the Mie scattering regime, the independent scattering approximation is valid up to volume fractions exceeding 10\% and often higher depending on scattering parameter, indicating the mean free path can usually be calculated much less expensively using the number density and the scattering cross-section of a single scatterer. We also study localization lengths and their dependence on particle loading.  In the case of heavy particles in a lighter matrix, we have been able to observe localization only at volume fractions $>30\%$ using the Landauer approach and only for high frequency modes, exceeding the vibration frequencies of the embedded nanoparticles. Using modal analysis we show that localization in nanoparticle laden materials is primarily due to energetic confinement rather than Anderson localization. Subsequentyly, we show that by using light particles in a heavy matrix the fraction of confined modes can be substantially increased.

\end{abstract}

\pacs{Valid PACS appear here}
\maketitle



\section{\label{sec:Introduction}Introduction}
%
Embedding nanoparticles into electronically functional materials is a known route to reduce thermal conductivity. Nanostructures provide additional scattering centers for phonons, thereby hindering thermal transport\cite{doi:10.1063/1.2188251}.  While this is an advantage in materials like thermoelectrics where reductions in the relative contribution of lattice thermal conductivity lead to direct increase in efficiency\cite{PhysRevLett.96.045901, doi:10.1021/nl080189t, doi:10.1021/nl8031982, doi:10.1063/1.4820151}, in other devices where thermal management is paramount, increased thermal resistance may be viewed as detrimental to performance. Despite the technological importance of embedded nanoparticles, there are many open questions about how to control thermal wavelength phonon-nanoparticle interactions.  

In many works related to modeling phonon-nanostructure interaction and associated thermal conductivity reduction, Boltzmann transport theory approaches are employed.  Typically a model of scattering cross section for an isolated phonon-nanoparticle interaction is first calculated and then used to estimate the relaxation time for the loaded case using number density as scaling factor\cite{PhysRevLett.96.045901, doi:10.1021/nl080189t, doi:10.1021/nl8031982, doi:10.1063/1.4820151, Feser:MieScattering}.  These approaches primarily differ in the level of sophistication used to calculate the scattering cross section, for example whether they are based on patching of limiting wave parameter cases\cite{doi:10.1021/nl8031982}, whether they can model multiple polarizations of phonons\cite{Feser:MieScattering}, whether they are based on analytic elastic continuum theory or an atomistic computational model of the scattering event\cite{PhysRevB.77.094302, PhysRevB.84.125426}, and whether they account for polydispersity\cite{doi:10.1063/1.2188251}.  It is sometimes taken for granted that the independent scattering approximation carries over for multiple nanoparticles, and that Boltzmann transport theory is a good description of the transport physics.  These assumptions are both questionable when the volume fraction of nanoparticles is sufficiently large.  In particular, effects like multiple and dependent scattering are ignored as well as the ability of disorder to induce different characteristics of the vibration modes such as `diffusons' and `locons' (localized modes)\cite{doi:10.1063/1.4955420,doi:10.1080/13642819908223054}.  Prasher\cite{10.1115/1.2194036, doi:10.1063/1.2794703} has analytically studied the effects of multiple and dependent scattering on phonon transport characteristics such as density of states, group velocity, and mean free path, for scattering and simultaneously scattering/absorbing random nanoparticulates, respectively. In a medium of randomly placed scatterers, the validity of the independent scattering approximation becomes uncertain at high volume fraction of scatterers, when the distance between the scatterers approaches the magnitude of the scattering wavelength, and the random nature of the arrangement of the scatterers is lost. The resulting spatial correlation leads to a phase correlation, and therefore, to near and far field interference between the scattered waves, and dependent scattering that generally increases the mean free path.

Localization of vibrational modes and the subsequent non-propagation of energy is another important aspect of disordered systems that arise in non-dilute nanoparticle-in-matrix composite materials.  Localized modes in other disordered systems such as amorphous materials\cite{PhysRevB.48.12589, PhysRevB.48.12581, doi:10.1080/13642819908223054, doi:10.1063/1.4955420, PhysRevB_McGaughey_2014}, alloys\cite{Seyf2017}, and random multilayers\cite{PhysRevB.103.045304, Ma_2020} have been previously studied using a variety of approaches, but to our knowledge just a few have concerned randomly embedded nanoparticles.  There are several approaches available to identify localized modes and measure their impact on transport.   Identification of localized modes using participation ratio (PR) is a simple option\cite{doi:10.1080/13642819908223054}, but one that requires relatively small atomistic systems (typically a $<5$nm cube in 3D) due to the poor scalability associated with solving the eigenvalue problem; when the modal decomposition is available, the localized modes can often be fit for their localization length as well\cite{doi:10.1080/13642819908223054}.  The Landauer approach is an alternative where mode-by-mode transmission functions are computed between two contacts/reservoirs and information from the length dependence of the transmission function can be used to determine the mean free path or localization length depending on the shape of the dependence\cite{PhysRevLett.101.165502, datta1997electronic}. Briefly, in the diffusive regime, $T_E\propto 1/L$, where $T_E$ if the transmission coefficient or transmission function and $L$ is the length of the domain.  In contrast, Anderson localized modes as well as interfacial/evanescent modes decay like $T_E\propto \exp(-L/\xi)$, which allows one to potentially fit both mean free path and localization length from transmission data (details discussed further in Sec. ~\ref{sec:Method}).  Savi\'{c} has used this approach enabled by the atomistic Green's function method (AGF) to study the localization length and it's effect on transport in isotope disordered carbon and boron nitride nanotubes\cite{PhysRevLett.101.165502}, a quasi-1D system.  Multiple scattering effects were clearly observed in simulations but with the caveat that phonon localization did not substantially alter the thermal conductivity compared to an approximate solution that treated only diffusive scattering regardless of the temperature. Hu\cite{PhysRevB.103.045304} has used a Green's function approach to study random multilayers of silicon-germanium with short period and has observed that for large disorder and short device lengths, the localized contribution to heat transport can be larger than the diffusive contribution.  Luckyanova\cite{doi:10.1126/sciadv.aat9460} used a similar approach to explore localized transport in roughened 3nm AlAs/3nm GaAs superlattices with random nanoparticles placed at the interfaces; we would note that although the simulations were nominally 3D, due to computational limitations the aspect ratio of the transmission domains were quite slender (1.68nm in each transverse direction by 1600nm in total length with one nanoparticle per interface and 6nm superlattice spacing), such that we believe this simulation represents quasi-1D localization.  At low temperature, Luckyanova observes a sharp reduction in conductivity when adding nanoparticles, which they attribute to phonon localization.  Ma\cite{Ma_2020} has used spectral decomposition within molecular dynamics to study phonon localization in 2D and 3D random multilayers of boron nitride/graphene.  Their approach intrinsically included anharmonicity.  They found that while 2D constrained RMLs had many localized modes, once flexural modes were introduced in 3D, anharmonic effects break phase often enough that even modes with low participation ratio transmit diffusively.    Note that in the prior works above, even if the simulations were performed nominally in 2D or 3D, the disorder itself was distributed in a quasi-1D manner.

In this work, we investigate how phonon transport is influenced by dense 2D randomly embedded nanoparticles with the goals of (1) understanding the limits of applicability of the individual scattering approximation and the eventual transition to multiple scattering, as well as (2) the observation of phonon localization and the characterization of localization length and it's dependence on characteristics such as density contrast and particle number density.   Working in 2D presents computational significant advantages over working in 3D.  Importantly, it becomes tractable to have domains that contain hundreds of nanoparticles in the modal decomposition approaches we use, and tens of thousands of nanoparticles in the Landauer approach used in this manuscript.  Thus we are able to have enough particles in the system to have both large transverse and horizontal extents and to therefor escape quasi 1D behavior.
\section{\label{sec:Method}Method}
\begin{figure}
\begin{center}
\includegraphics[width=0.45\textwidth]{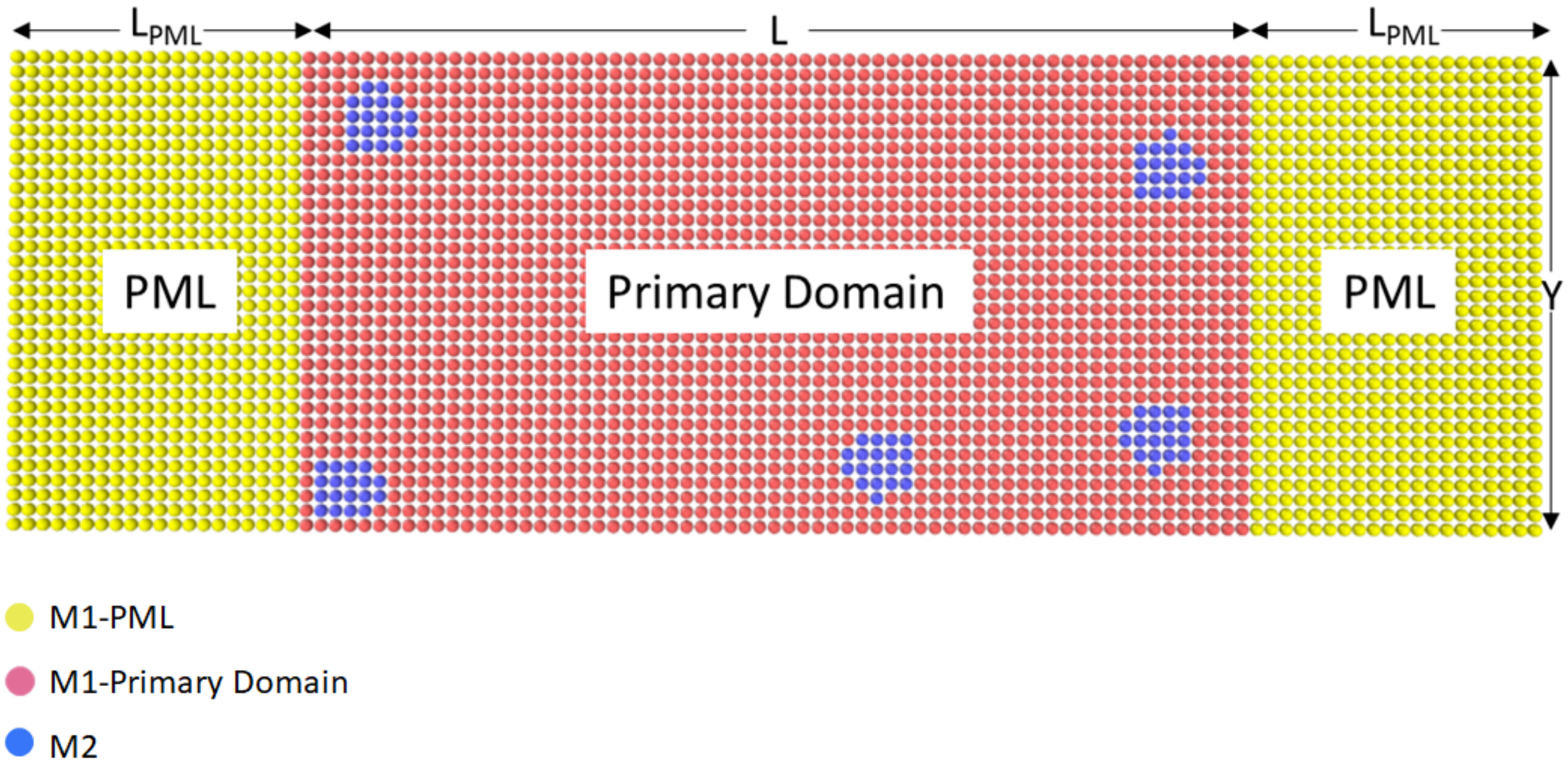}
\end{center}
\caption{\label{fig:fdpml_TE} Schematic diagram of the domain used for calculating transmission coefficient. }
\end{figure}

A Landauer approach is used to computationally characterize phonon scattering and localization in 2D domains having randomly positioned particulate scatterers, with the mode-by-mode transmission coefficients calculated by the frequency domain perfectly matched layer (FDPML) method\cite{doi:10.1063/1.4929780}. We have previously described the FDPML method in the context of 1D and 3D transmission coefficient calculations \cite{doi:10.1063/1.4929780, PhysRevB.95.125434}, as well as calculation of scattering cross-section in 2D and 3D.  Briefly, the transmission coefficient calculation works in the following manner:  atoms are modeled as bonded by harmonic interatomic force constants, which allows the equations of motion to be written at any particular frequency as a linear system of algebraic equations in the frequency domain.  A scattering system of interest called the primary domain (Fig.~\ref{fig:fdpml_TE}, red and blue) is connected to left and right contact regions called perfectly matched layers (PMLs; Fig.~\ref{fig:fdpml_TE}, yellow); the equations of motion in the perfectly matched layer are adjusted so that they serve as perfect absorbers while being nearly reflectionless for phonons arriving from the primary domain.  In effect, the PMLs are designed so that the primary domain cannot tell the difference between being connected to infinitely extended non-absorbing contacts and the PML.  Then, a plane incident phonon with wavevector, k, and polarization, p, with associated frequency, $\omega(k)$, is simulated as being emitted from the left PML into the primary domain.  The frequency domain response is then computed by solving the system of algebraic equations for the displacements of all the atoms in the frequency domain.   The displacements in the left and right PML can be related to the total reflected and transmitted energies into the primary domain, respectively.  Details on the method and design of the PML are available in Refs. \citenum{doi:10.1063/1.4929780} and \citenum{PhysRevB.95.125434}. 
%

The primary domain in this work (Fig.~\ref{fig:fdpml_TE}) has longitudinal length L and transverse period length Y (the y-direction is periodic), which consists of a square lattice of matrix atoms with mass $M_1$ connected by nearest and 2nd-nearest neighbor interatomic force constants corresponding to linear springs having spring constants $G_1$ and $G_2$.  We embed randomly placed 2D nanocylinders with atoms of mass $M_2$, having diameter, $D$, and volume fraction, $V_f$.  Random placements of the embedded cylinders are artificially generated using a hard cylinder molecular dynamics algorithm with random initial velocities and allowing $5N_{nc}$ particle-particle collisions ($N_{nc}$ = number of nanocylinders);  the cylinders in the MD model are thus intrinsically prevented from having any overlap.  After the last collision, any lattice positions which reside inside the cylinders, $||x-x_c||<D/2$ are labelled as having mass $M_2$.   Unless otherwise specified we use parameters in Table~\ref{tab:parameters}. The values of M1 and M2 are taken as 28 amu and 72 amu respectively, representing that of a Si matrix and Ge particles. The values of the spring constant $G_1$ and $G_2$ are chosen to give longitudinal speed of sound similar to Si and we choose $G_2=G_1/2$ to create isotropic properties in the long wavelength limit.  The same spring constants are used for both matrix and nanoparticulates.
 \begin{table}
 \caption{\label{tab:parameters}Parameters used in 2D model}
 \begin{ruledtabular}
 \begin{tabular}{cc}
 Parameter & Value\\
 \hline
$M_1$ & 72 a.m.u.\\
$M_2$ & 28 a.m.u.\\
$G_1$ & 21.35 N/m \\
$G_2$ & 10.7 N/m \\
$a$ & 0.273 nm \\
Polarization & Longitudinal\\
Direction of incident wave & [10]\\
D & 5a\\

 \end{tabular}
 \end{ruledtabular}
 \end{table}

The final system of equations solved by the FDPML algorithm is of the form,   
\begin{equation}\label{eq:FDPML}
    (\mathbf{M}\omega^2 + \mathbf{G} + \mathbf{M}\mathbf{\sigma^2}-2i\mathbf{M}\mathbf{\sigma})\mathbf{u}^{scat}=-(\mathbf{M}\omega^2 + \mathbf{G} )\mathbf{u}^{inc}
\end{equation}
where, $\mathbf{M}$, is a diagonal matrix containing the mass of each atom, $\mathbf{G}$ is a matrix containing the IFC information, and $\mathbf{\sigma}$ is a diagonal matrix containing the damping coefficients associated with each atom (in the primary domain these coefficients are all zero)\cite{doi:10.1063/1.4929780}. In Eq.~\ref{eq:FDPML}, the equations of motion are designed so that the damping does not affect the incident wave. As in Ref. \citenum{doi:10.1063/1.4929780}, we use damping coefficients, $\sigma$, that increase parabolically from zero at the interface of the PML and the primary domain to a maximum at the edge of the PML. For computational efficiency components of the damping coefficients are graded relative to their proximity to the edge of the PML as 
\begin{equation}
    \sigma_x = \sigma_y = \sigma_{max}\left(\frac{\ell-L_{PML}}{L_{PML}}\right)^2
\end{equation}
In Ref.\citenum{doi:10.1063/1.4929780}, we described a method based on dimensional analysis to determine suitable parameters $L_{PML}$ and $\sigma_{max}$ that balance the need for a small number of atoms in the PML and acceptable simulation accuracy. To ensure an accuracy of $\approx 10^{-4}$ for the transmission coefficient values, we keep the ratio of $L_{PML}/\lambda=8$. Thus, the value of $L_{PML}$ is increased for longer wavelengths, thus increasing the total number of atoms in the computational domain for which Eq.~\ref{eq:FDPML} must be solved. The total number of atoms for the largest domain studied in this work is approximately $10^6$, small enough to be readily solved using direct methods on shared memory computers. The value of $\sigma_{max}$ is similarly determined by keeping the ratio of $\sigma_{max}$ to the time taken for a wave to propagate $L_{PML}$, constant. 

\begin{figure}
\begin{center}
\includegraphics[width=0.45\textwidth]{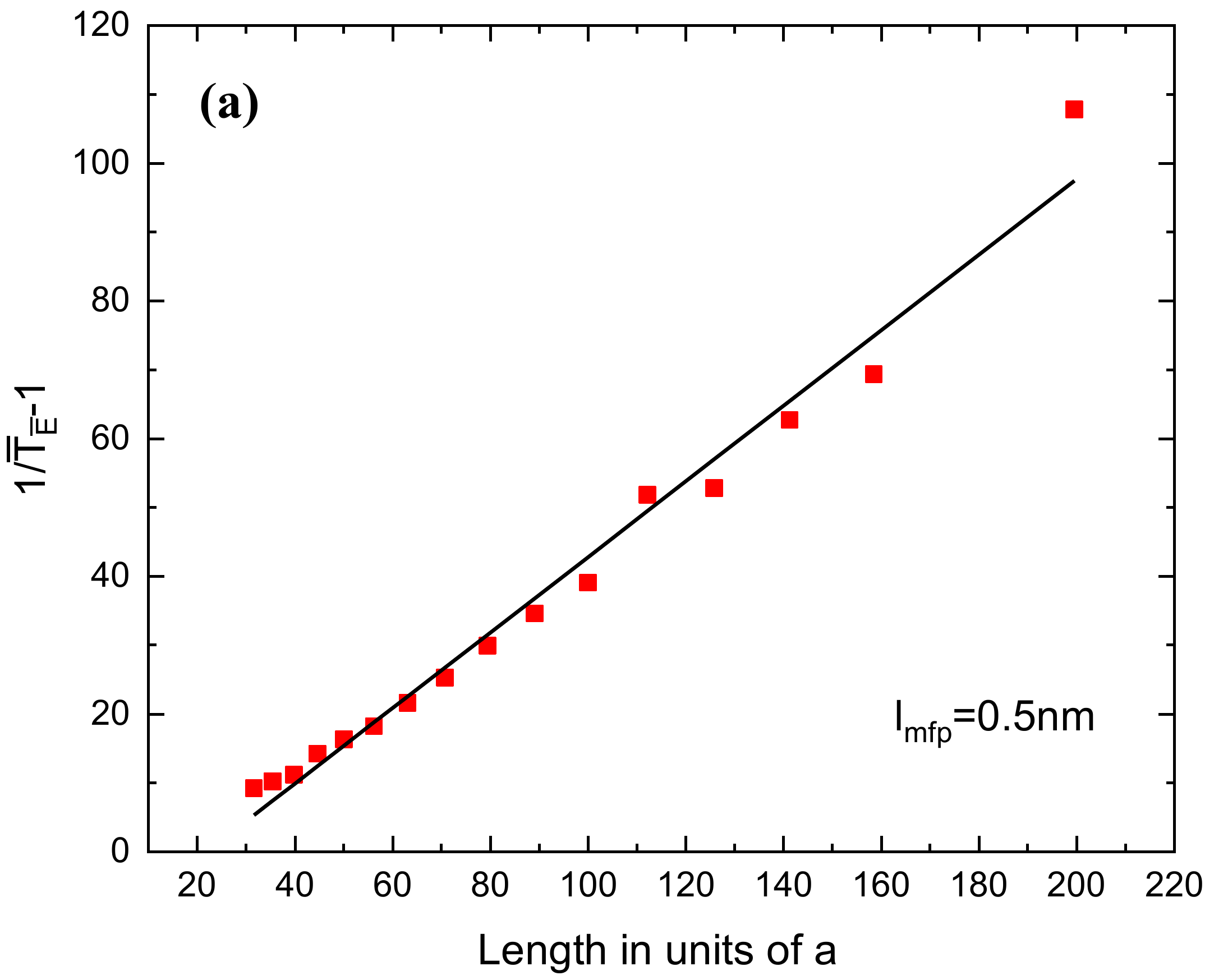}\\
\includegraphics[width=0.45\textwidth]{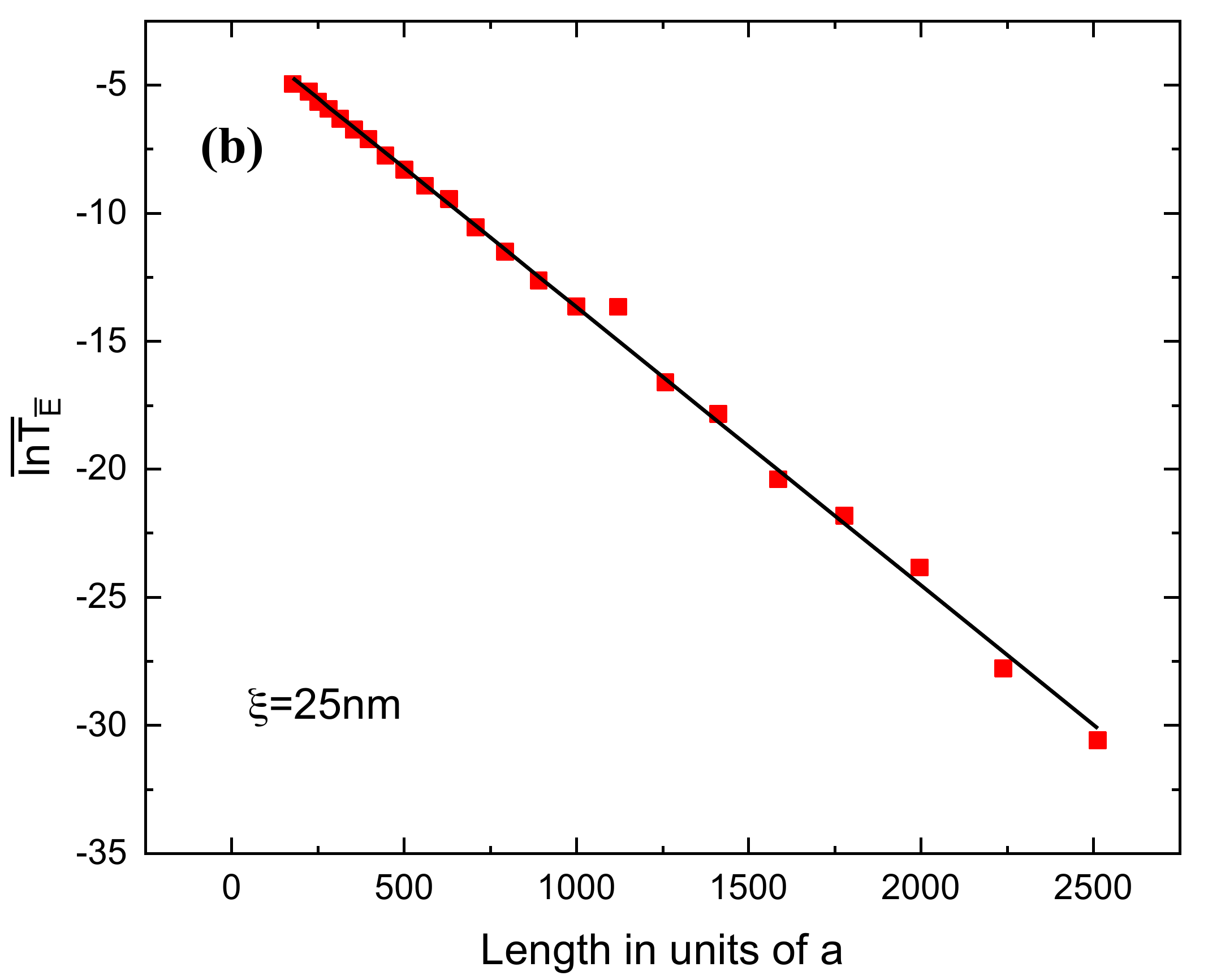}
\end{center}
\caption{\label{fig:mfp_loc_fits} (a). $\ell_{mfp}$ is obtained from a fit to $1/\overline{T_E}-1$ versus $L$. (b) $\xi$ is obtained from a fit to $\overline{\ln {T_E}}$ versus $L$. The parameters used to obtain this figure are given in Table.1 and the transverse Y length is equal to 128a.}
\end{figure}

After solving for the displacements, the energy absorbed by each atom in the PML is calculated from the magnitude of the scattered displacement of that atom.  The total amount of energy absorbed in each PML is the sum of the energy absorbed by all of the atoms in that PML\cite{doi:10.1063/1.4929780}, 
\begin{equation}
    Q_{PML}=\omega^2 \sum_{n=1}^{n_{PML}}{m_n(\sigma_{x,n}|u_n^{x,scat}|^2+\sigma_{y,n}|u_n^{y,scat}|^2)}
\end{equation}
We calculate the transmission coefficient for an incident wave travelling in the positive x-direction from the ratio of the amount of energy absorbed in the right PML to the total amount of energy absorbed, which is the sum of the amount of energy absorbed in the left PML and the right PML,
\begin{equation}
    T_E = \frac{Q_{PML,right}}{Q_{PML,left}+Q_{PML,right}}
\end{equation}
The transmission coefficients, $T_E$, computed by the FDPML method are then used to obtain the mean free path, $\ell_{mfp}$, and the localization length, $\xi$, of the disordered systems by identifying the different scaling trends of transmission with length in the diffusive and the localized regimes.  In particular, in the ballistic-to-diffusive regime, it is well known that the transmission follows\cite{datta1997electronic} $T_E(L)=1/(1+L/\ell_{mfp})$, and for localized modes transmission follows, $T_E(L)=\exp(-L/\xi)$. If the localization length and mean-free-path are well separated ($\xi\gg\ell_{mfp}$), then it is possible to extract both the mean free path and the localization length associated with the scattering domain.  In practice, we perform fits of the configurational average $\overline{T_E(L)}=\frac{1}{N_{config}} \sum_{n=1}^{N_{config}}T_{E,n}(L)$, similar to Refs. \citenum{PhysRevB.74.121406} and \citenum{PhysRevLett.101.165502}, and also follow their scheme to determine the shift from the diffusive to the localized regime by using statistical criterions. $T_E$ values are calculated for 100 different configurations of the primary domain having the randomly embedded nanoparticles and these calculations are performed at each length, for a range of lengths up to $L=2512a$ ($a$ is the lattice spacing), while keeping the number density of the nanoparticles, $\eta$, and $Y$ constant.  $\ell_{mfp}$ is obtained by rearranging the linear form to $1/\overline{T_E}-1=L/\ell_{mfp}$  and performing a linear regression.  However, as the length increases the localized regime is approached, where $T_E=1/(1+L/\ell_{mfp})$ becomes a poor fit as the transmission no longer scales inversely with length once the localization length is approached. For $L>L_c$, the localization length, $\xi$, is then extracted by fitting $\overline{\ln{T_E}}=-L/\xi$. To determine the crossover from the ballistic/diffusive to the localized regime, $L_c$, we use the statistical criterion\cite{PhysRevB.74.121406} that transmission coefficients that satisfy $\Delta T_E/\overline{T_E} \leq 1$, where $\Delta T_E\equiv \sqrt{\overline{T_E^2}-\overline{T_E}^2}$ are used to obtain $\ell_{mfp}$, and transmission coefficients that satisfy $\Delta T_E/\overline{T_E} > 1$ and $\Delta \ln{T_E}/\overline{\ln{T_E}}\leq 1$, where $\Delta \ln{T_E} \equiv \sqrt{\overline{\ln{T_E}^2}-\overline{\ln{T_E}}^2}$, are used to extract $\xi$. The plots of Fig.~\ref{fig:mfp_loc_fits}(a) and (b) are obtained when a longitudinal wave of $\lambda=5a$ is incident on a domain consisting of $\approx48\%$ by volume fraction of embedded nanocylinders. The characteristic lengths are each obtained from the reciprocal of the slope of the fits. 

For a truly 2D simulation, results should be independent of the transverse length, $Y$ (Figure~\ref{fig:fdpml_TE}).   Thus, reported mean free paths, $\ell_{mfp},$ and the localization lengths, $\xi$, were calculated by increasing Y until the $\ell_{mfp}$, and $\xi$ becomes approximately independent of Y and then taking the mean of those $\ell_{mfp}$ and $\xi$ points that have approximately converged. The plots of $\ell_{mfp}$ versus $Y$, and that of $\xi$ versus $Y$, are provided in the supplementary materials\cite{supplemental_info}.

\section{Results}

\subsection{Variation of mean free path with particle number density}
\begin{figure}
\begin{center}
\includegraphics[width=0.45\textwidth]{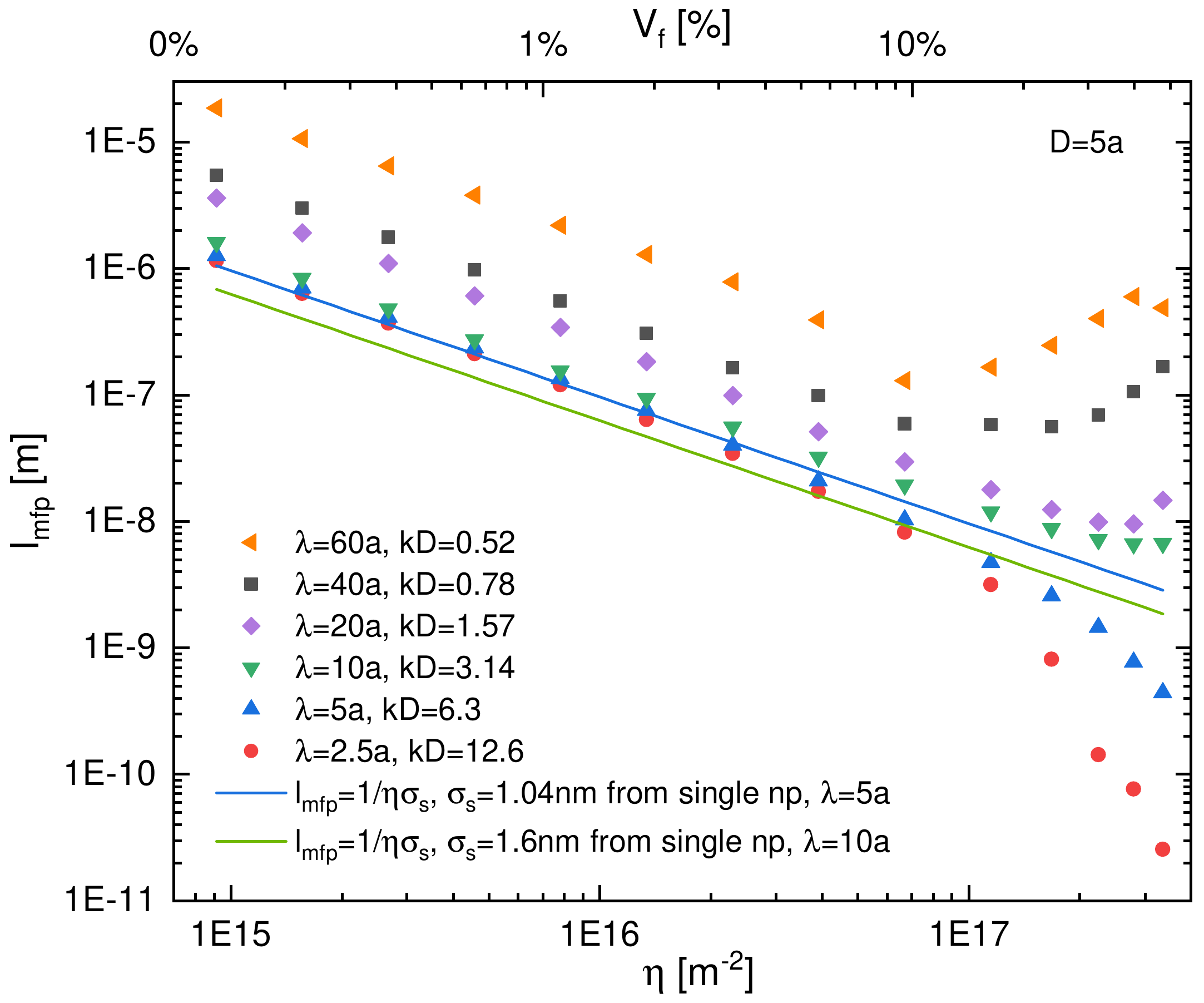}
\end{center}
\caption{\label{fig:mft_nd}$l_{mfp}$ versus $\eta$ for waves scattering with particles of diameter, $D=5a$. The dotted plots for $\lambda= 2.5a, 5a, 10a, 20a, 40a,$ and $60a$, are obtained from domains containing random nanoparticles, and the line plots for $\lambda= 5a, 10a$ are obtained from the equation, $l_{mfp}=1/\sigma_s\eta$ where $\sigma_s$ is the scattering cross section of a single nanoparticle of $D=5a$. $kD$ is the scattering parameter, where $k$ is the wavenumber given by $k=2\pi/\lambda.$}
\end{figure}

When phonons scatter independently with nanoparticles, the mean free path, $\ell_{mfp}$,  scales inversely with the product of the scattering cross section, $\sigma_s$ and number density, $\eta$, of the nanoparticles, e.g.  $\ell_{mfp}=1/(\sigma_s \eta)$. We use the Landauer approach to calculate $\ell_{mfp}$ for a range of $\eta$ of randomly embedded nanoparticles, corresponding to volume fractions, $V_f$, ranging from 0.07\% to 48\%. Fig.~\ref{fig:mft_nd} shows a log-log plot of $\ell_{mfp}$ versus $\eta$ for a variety of incident wavelength varying from $\lambda=2.5a$  to $\lambda=60a$, for a fixed scatterer diameter D=5a. A major result of this work can be seen in Fig.~\ref{fig:mft_nd}, where the independent scattering approximation $\ell_{mfp}\approx(\sigma_s \eta)^{-1}$ is shown to hold for a wide range of wavelengths up to $V_f>10\%$ before it begins to deviate depending on relative wavelength. The worst case deviation begins in the Rayleigh regime, corresponding to the onset of dependent scattering effects; for higher volume fractions, the mean free path’s dependence of volume fraction weakens and eventually begins to increase with volume fraction.  Prasher has analytically studied multiple and dependent scattering in 3D for compressive waves in this regime with similar conclusions (see Fig. 7 of Ref. \citenum{10.1115/1.2194036}).  For the Mie Scattering and geometric scattering regimes, the independent scattering approximation is observed to hold to much higher volume fractions. For example, Fig.~\ref{fig:mft_nd} shows that $\ell_{mfp}$ continues to scale inversely with $\eta$ for volume fractions past $V_f>30\%$ in the cases ($\lambda=10a, 20a$), with the minimum $\ell_{mfp}$ being reached at increasing larger volume fractions for smaller wavelength.   Previously, it has been shown that the size of nanoparticles which most effectively reduce thermal conductivity (at least for monodisperse systems) are those in the Mie regime.

For comparison we have also included a plot of $(\sigma_s \eta)^{-1}$,  where the value of $\sigma_s$ used is the scattering cross section for a single nanoparticle of the same diameter $D=5a$ with $\lambda=5a,10a$. The scattering cross section is also calculated by the FDPML method, using the method in Ref. \citenum{doi:10.1063/1.4929780}.  In this variation of the method, a single nanoparticle is located at the center of the primary domain which measures $128a\times128a$, with a PML surrounding the primary domain on all sides, rather than just the left and right. The “incident” wave is applied to all the atoms in the primary and PML domains, except for the atoms constituting the nanoparticle. All the other parameters used in the calculation of the scattering cross section are the same as those given in Table 1. We calculate scattering cross section, $\sigma_s$, from the ratio of the rate of scattered wave energy absorbed by the PML, as given by Eq. (3) to the incident wave energy flux. In a 2D square lattice, with a single atom per unit cell and unit incident wave amplitude, the incident flux is given by $(m\omega^2 v_g)/a^2$, where $m$ is the mass of the atoms on which the incident wave is applied, and $v_g$ is the group velocity. The results are shown as the lines in Fig.~\ref{fig:mft_nd} as solid blue and green lines.  Although the $\ell_{mfp}\propto\eta^{-1}$ scaling is observed in both cases, there appears to be a slight underprediction of mean free path values in the $\lambda=5a$ case, and a larger discrepancy for $\lambda=10a$ case, when using the single nanoparticle scattering cross section compared to those observed from the large multiparticle simulations. 

\begin{figure}
\begin{center}
\includegraphics[width=0.35\textwidth]{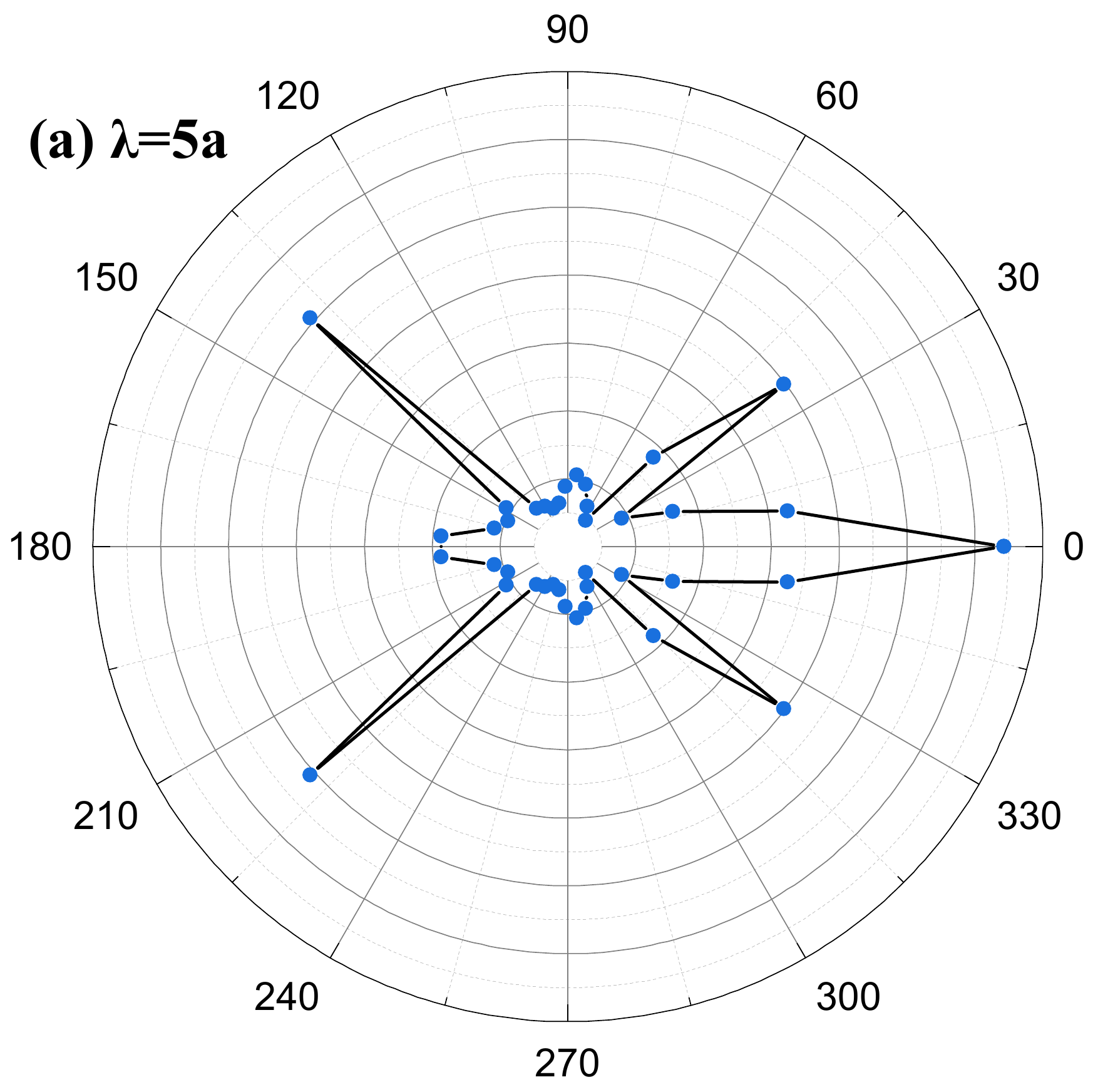}\\
\includegraphics[width=0.35\textwidth]{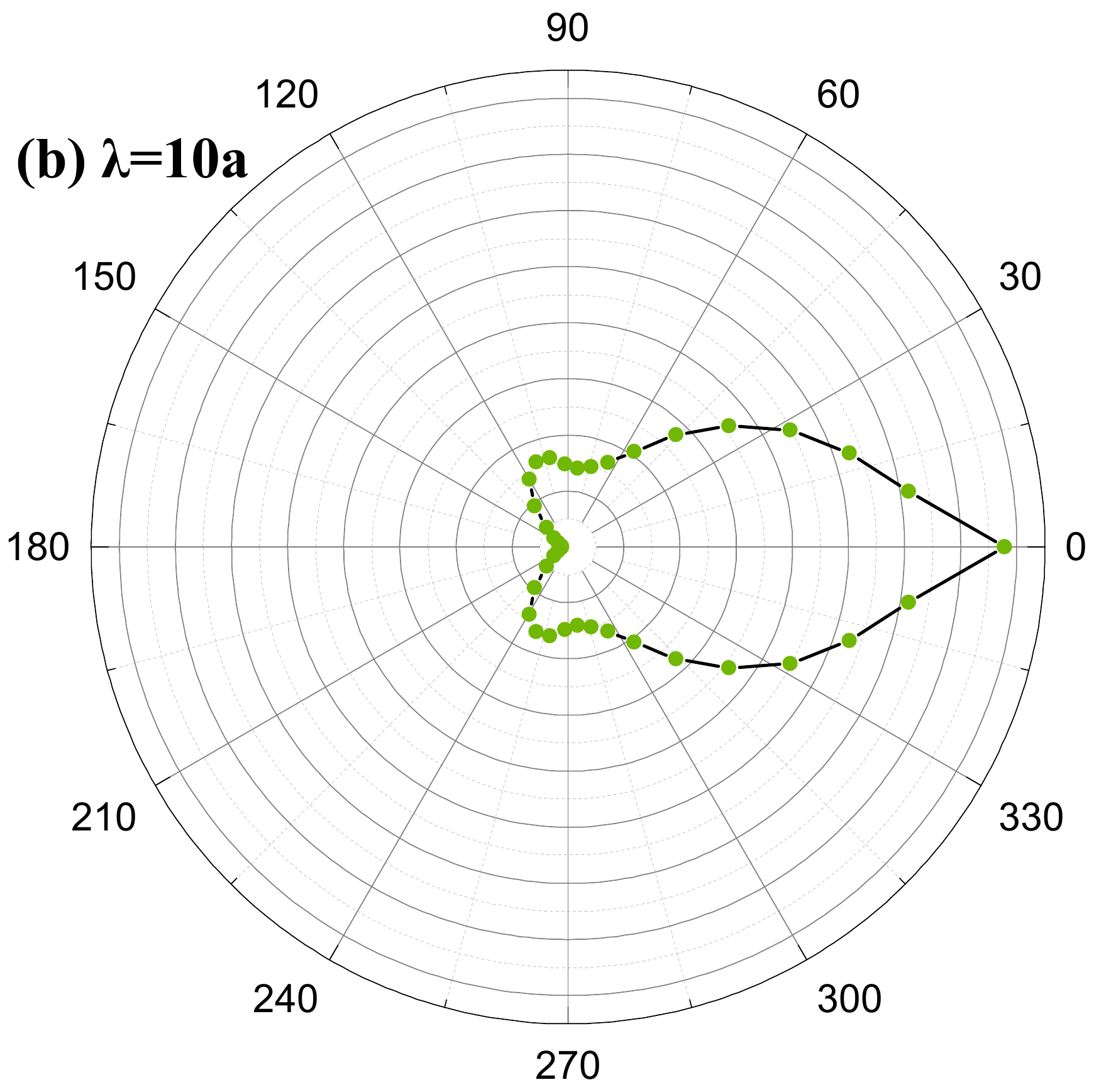}
\end{center}
\caption{\label{fig:pf} (a). Scattering phase function for diameter, $D=5a$, LA incident wave along $[10]$ with (a) $\lambda=5a$, and (b) $\lambda=10a$.}
\end{figure}

We have tested several hypotheses that could explain the discrepancy. One possibility is related to the directionality of phonon scattering. The expression, $\ell_{mfp}=1/(\sigma_s \eta)$ assumes that energy is scattered isotropically; but in fact, in our multiparticle study the transmission coefficient is calculated from the amount of energy that is scattered forward through the domain. Scattering of phonons from nanoparticles is known to be directionally dependent\cite{PhysRevB.77.094302, 10.1115/1.1622718}, the degree of which varies with factors such as polarization, wavelength, size, and mass ratio. Thus, a bias toward forward scattered energy will result in larger transmission coefficient values and thus larger $\ell_{mfp}$, than that obtained from the single nanoparticle scattering cross-section FDPML calculation, which accounts for total energy scattered by the particle without regard for direction. We have made a modification to the FDPML method which allows the quantification of directional energy scattering from a nanoparticle. based on the spatial dependence of absorbed energy in the PML (methodology described in the Supplmental Materials\cite{supplemental_info}), and provides equivalent information to the MD approach by Zuckermann\cite{PhysRevB.77.094302}.  Fig.~\ref{fig:pf}a shows that the scattering is not isotropic but is only somewhat biased in the forward direction for $\lambda=5a$, leading to a small underprediction in mean free path (Fig.~\ref{fig:mft_nd}).   For $\lambda=10a$, the forward bias of scattering is much more extreme, leading to a larger discrepancy between the individual scattering limit and the multiparticle simulations.  This could therefore explain the discrepancy between individual scattering theory and multiparticle observations at low number densities in Fig.~\ref{fig:mft_nd}.

Another possibility that we have explored and been able to discount - purely numerical in nature - is that the particle sizes in the multiparticle simulation are smaller than the targeted $D=5a$ due to discretization approach. Our numerical approach of treating all atoms within radius $D/2$ of the center of a circle as atoms within the nanoparticle causes a different number of atoms to appear in each particle depending on where the center of the circle lies (i.e. the discrete particles are not exactly cylinders and their sizes are polydisperse, see Fig 1.). However, we have done a numerical investigation (more information in the Supporting Information\cite{supplemental_info}) to study this effect and find that the number of atoms in each particle is clustered tightly about the mean (mean: 19.6 atoms, std: 0.97 atoms) and the mean is close to the value used for the scattering cross-section calculation (21 atoms). Thus, the difference in particle sizes between the scattering cross-section calculation and the multiparticle simulation are not large enough to explain the differences observed for $\lambda=10a$ in Fig.~\ref{fig:mft_nd}.  

\subsection{Phonon Localization}
Localization length can be estimated using the Landauer approach whenever an exponential decay of transmission with domain length can be detected (Fig.~\ref{fig:mfp_loc_fits}b).  For the range of nanoparticle volume fractions (0.7\%-48\%), wavelengths ($\lambda=$60a, 40a, 20a, 10a, 5a, 2.5a), and domain lengths we have studied (up to $L=2512a$), we have not been able to detect any localization in most cases.  It was only clearly observed for incident waves of our smallest wavelength, $\lambda=2.5a$, and then at only the highest nanoparticle volume fractions of approximately 30\%-48\%.  As before, we have confirmed the $Y$-independence of $\xi$ in these cases (see Supplemental Materials\cite{supplemental_info}).  Fig.~\ref{fig:loc_nd} shows $\xi$ decreases very rapidly with increasing $\eta$ in the case $D=5a$, $\lambda=2.5a$ (longitudinal).  

\begin{figure}
\begin{center}
\includegraphics[width=0.45\textwidth]{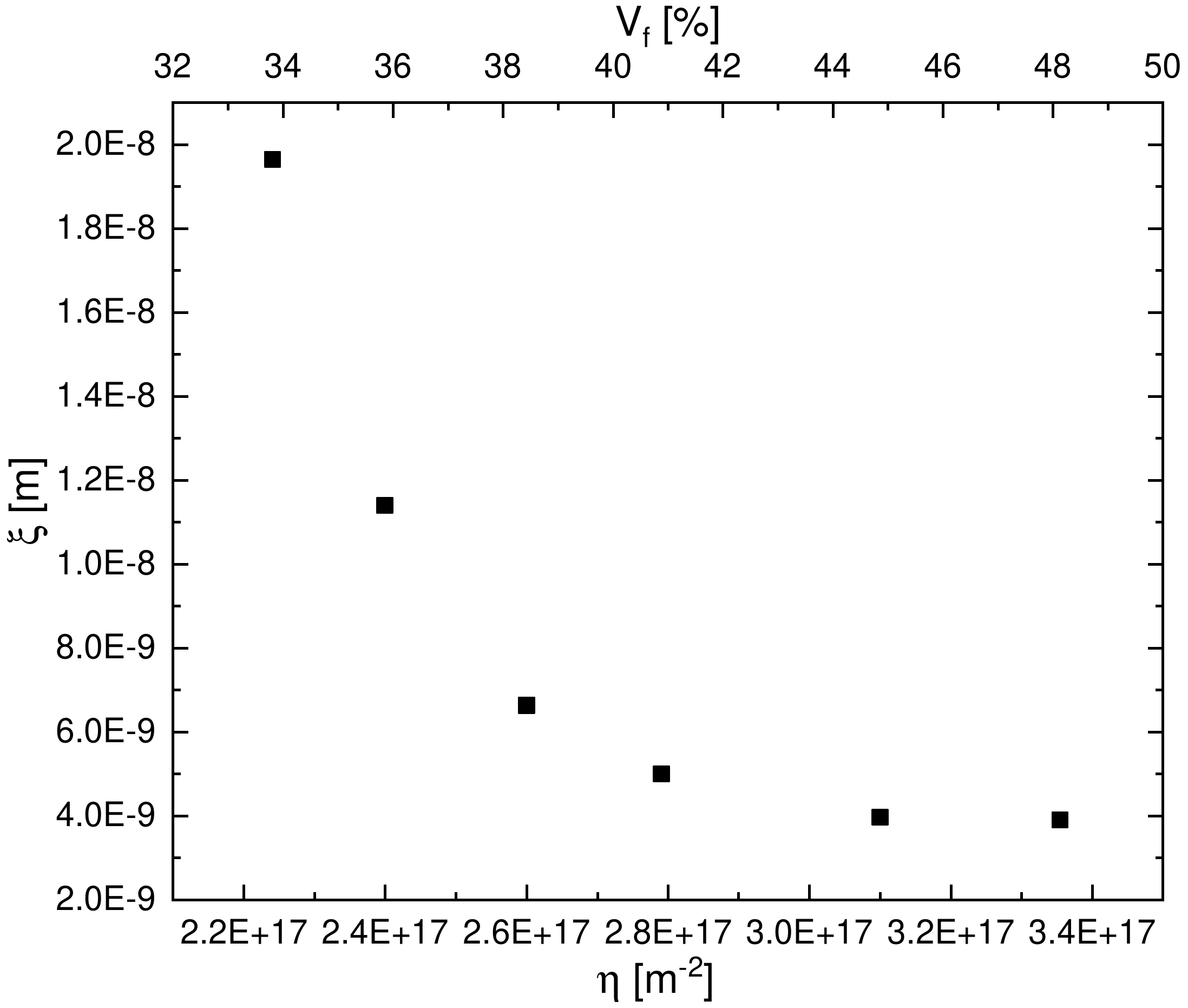}
\end{center}
\caption{\label{fig:loc_nd} Localization length as a function of number density/volume fraction for wavelength, $\lambda=2.5a$, diameter, $D=5a$. Other parameters in Table I.}
\end{figure}

For comparison, we have also used a modal analysis approach to identify and characterize localized modes.  The approach is similar to that used by Allen et al.\cite{doi:10.1080/13642819908223054}. In this approach, the entire disordered domain is regarded as a large supercell, and one solves for the eigenvectors and eigenvalues of the dynamic matrix.  The domain is nearly identical to the primary domain of our FDPML method but instead of having PMLs at its left/right boundaries, it applies periodic boundaries on all sides of the domain.  The eigenvalues of the dynamic matrix give the square of the frequency while the eigenvectors give the mass-modified displacements of each atom in the supercell. 

\begin{figure}
\begin{center}
\includegraphics[width=0.45\textwidth]{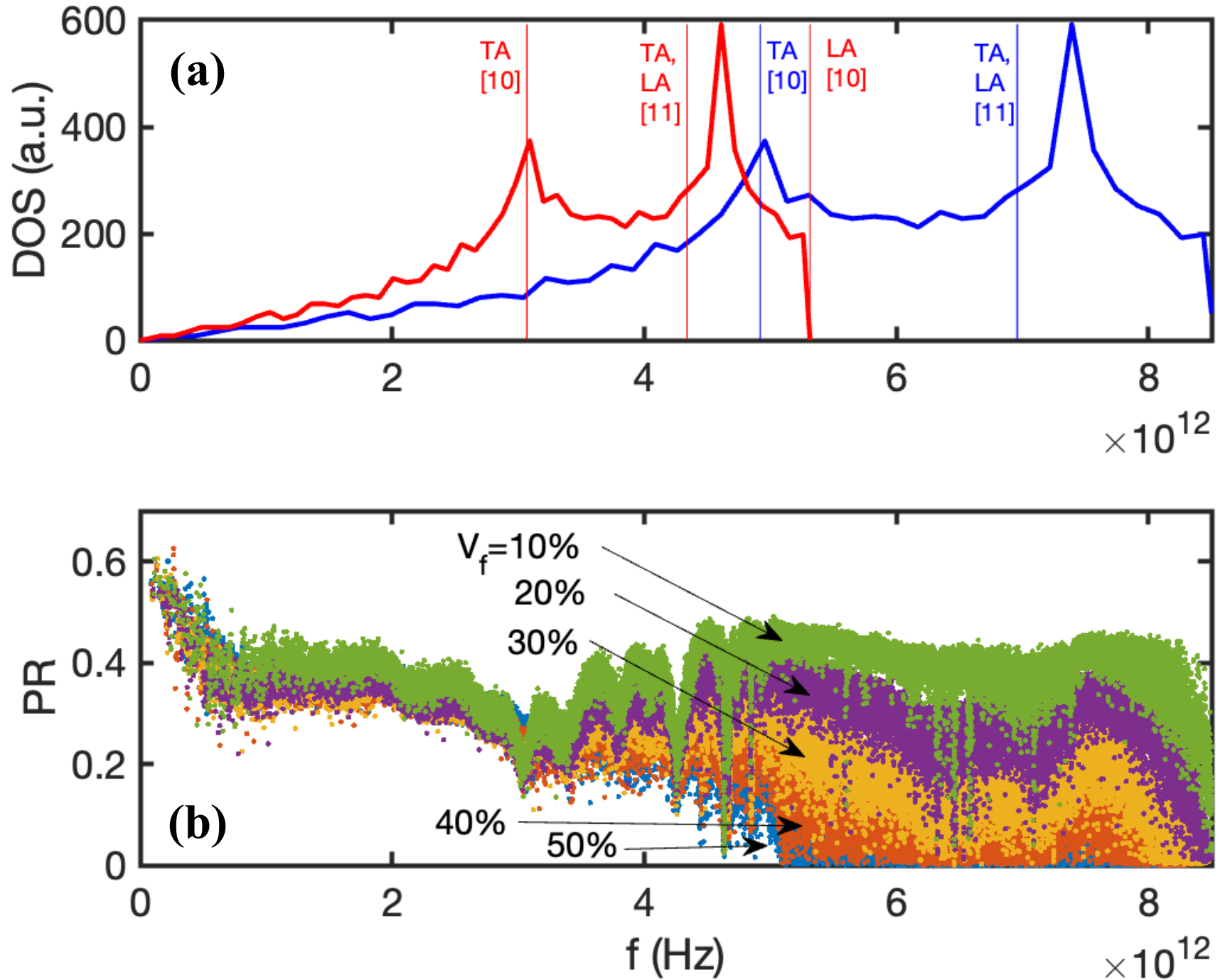}
\end{center}
\caption{\label{fig:DOSandPR} (a) Density of states for the pure materials with mass $M_1 =28 m_p$ (blue), and $M_2=72 m_p$ (red).  Frequency cutoffs at high symmetry points in the first Brillouin zone are shown a vertical lines.  (b)  Participation Ratio vs frequency for various volume fractions.}
\end{figure}

The size of the dynamic matrix is $2 N_{tot}\times 2N_{tot}$ for a domain with $N_{tot}$ atoms in 2D, and thus solving the eigenvalue problem gives $2N_{tot}$ eigenmodes.
The participation ratio ($PR$) of a mode can be used to determine whether a mode is localized or delocalized.  $PR$ of the $m$-th mode is given by \cite{doi:10.1063/1.4955420}, 
\begin{equation}
    PR=\frac{(\sum_n \epsilon_{m,n}^2)^2}{N_{tot}\sum_n \epsilon_{m,n}^4}
\end{equation}
Where $\epsilon_{m,n}$  is the nth degree of freedom of the eigenvector associated with the $m$th mode. The definition is designed such that if all degrees of freedom vibrate with equal amplitude and the eigenvector has been normalized, then the PR approaches unity.  If only a small number of atoms have significant displacements, then PR is small ($PR\ll 1$) and the mode can be considered localized\cite{doi:10.1063/1.4955420,doi:10.1080/15567265.2018.1519004}.  In the limiting case, $PR=1/N_{tot}$ for just one degree-of-freedom moving.  The exact threshold for localization is subjective, but we use a cutoff of $PR<0.1$.  Fig.~\ref{fig:DOSandPR}b plots $PR$ for modes in a $128a \times 128a$ domain of randomly embedded nanoparticles ($D=5a$, other properties from Table 1) for increasing volume fraction.  Fig.~\ref{fig:DOSandPR}a shows the corresponding density of vibrational states for the constituent matrix (blue) and nanoparticle materials (red).  Comparing Fig.~\ref{fig:DOSandPR}a and~\ref{fig:DOSandPR}b, it becomes clear that localized modes generally form beyond the upper band edges of the heavier material (in this case the nanoparticle material).  This suggests that the localization seen is primarily a confinement effect restricted to high frequency modes rather than an interference effect due to multiple scattering (i.e. Anderson localization); in other words, the localized modes are caused by the fact that modes cannot propagate through the heavy phase above $\approx$5.3~THz and therefore if any region of light atoms happen to be surrounded (or nearly surrounded) by heavier ones, modes can exist in that region that cannot in the surrounding material.   In Fig.~\ref{fig:displacement_patterned_heavyparticle}, some example displacement patterns are shown for localized modes at various frequencies, and they confirm this hypothesis.

\begin{figure}
\begin{center}
\includegraphics[width=0.45\textwidth]{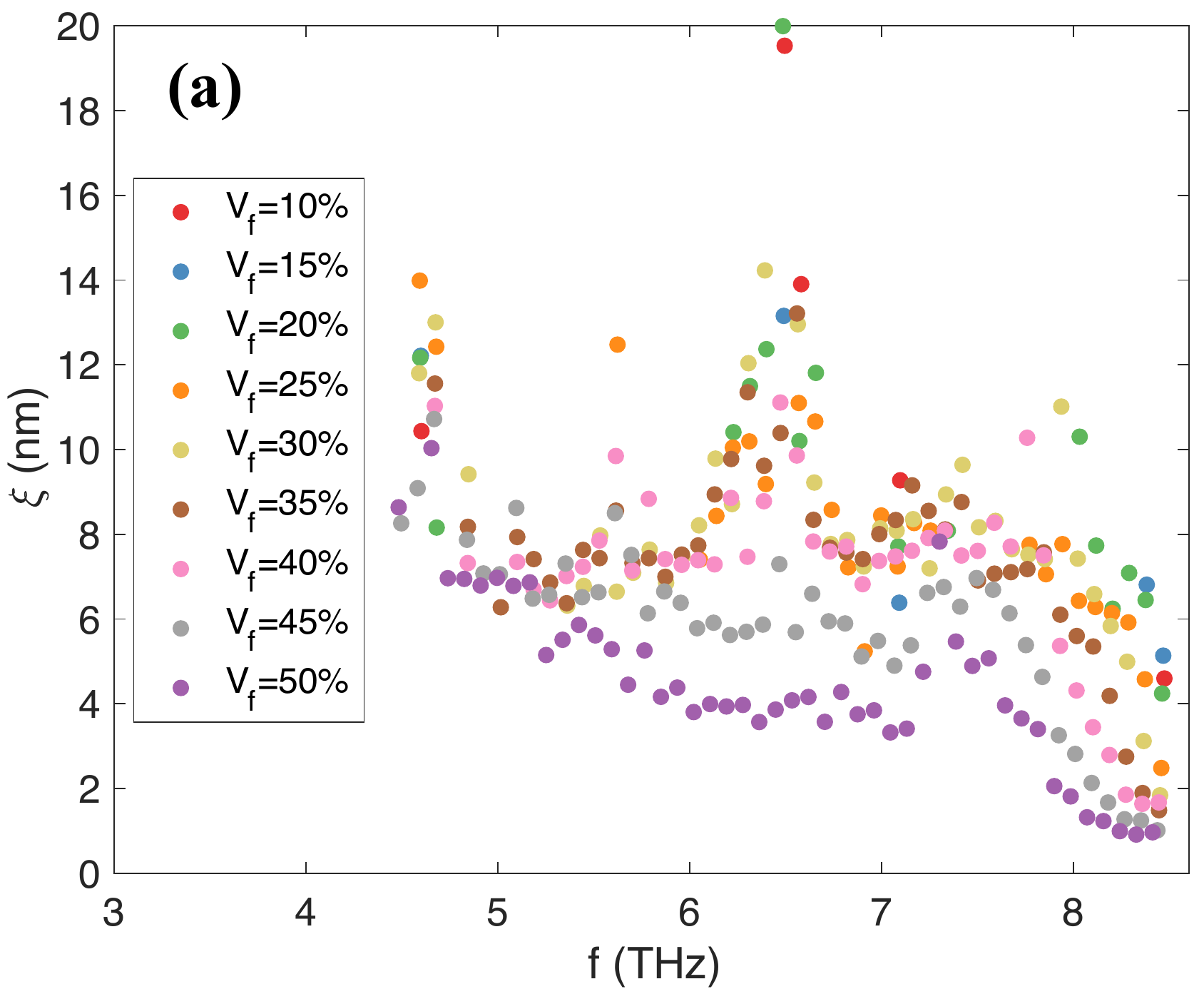}\\
\includegraphics[width=0.45\textwidth]{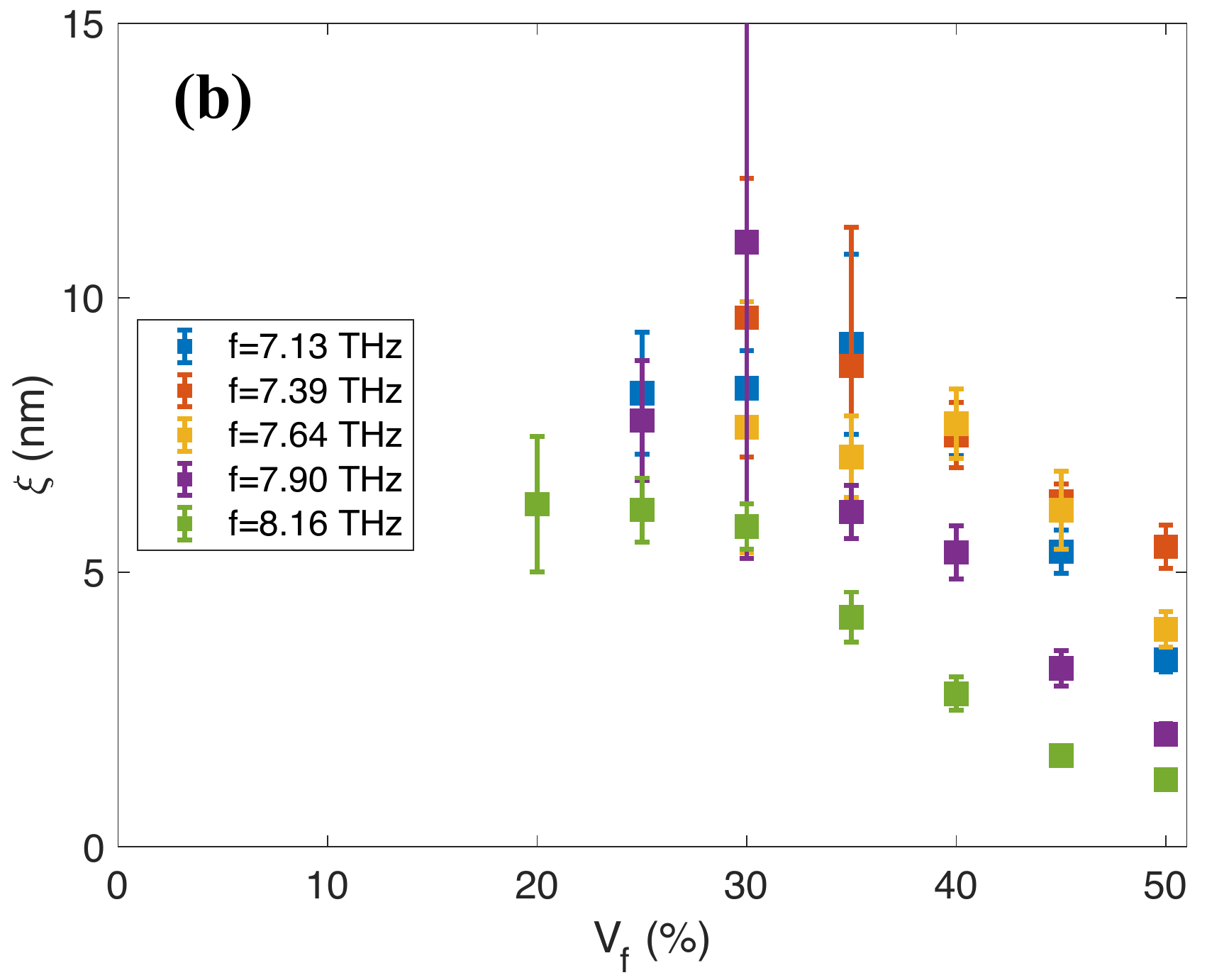}
\end{center}
\caption{\label{fig:localization_length_modal_decomp} Average localization lengths at various (a) frequencies and (b) volume fractions.}
\end{figure}

\begin{figure*}
\begin{center}
\includegraphics[width=.4\textwidth]{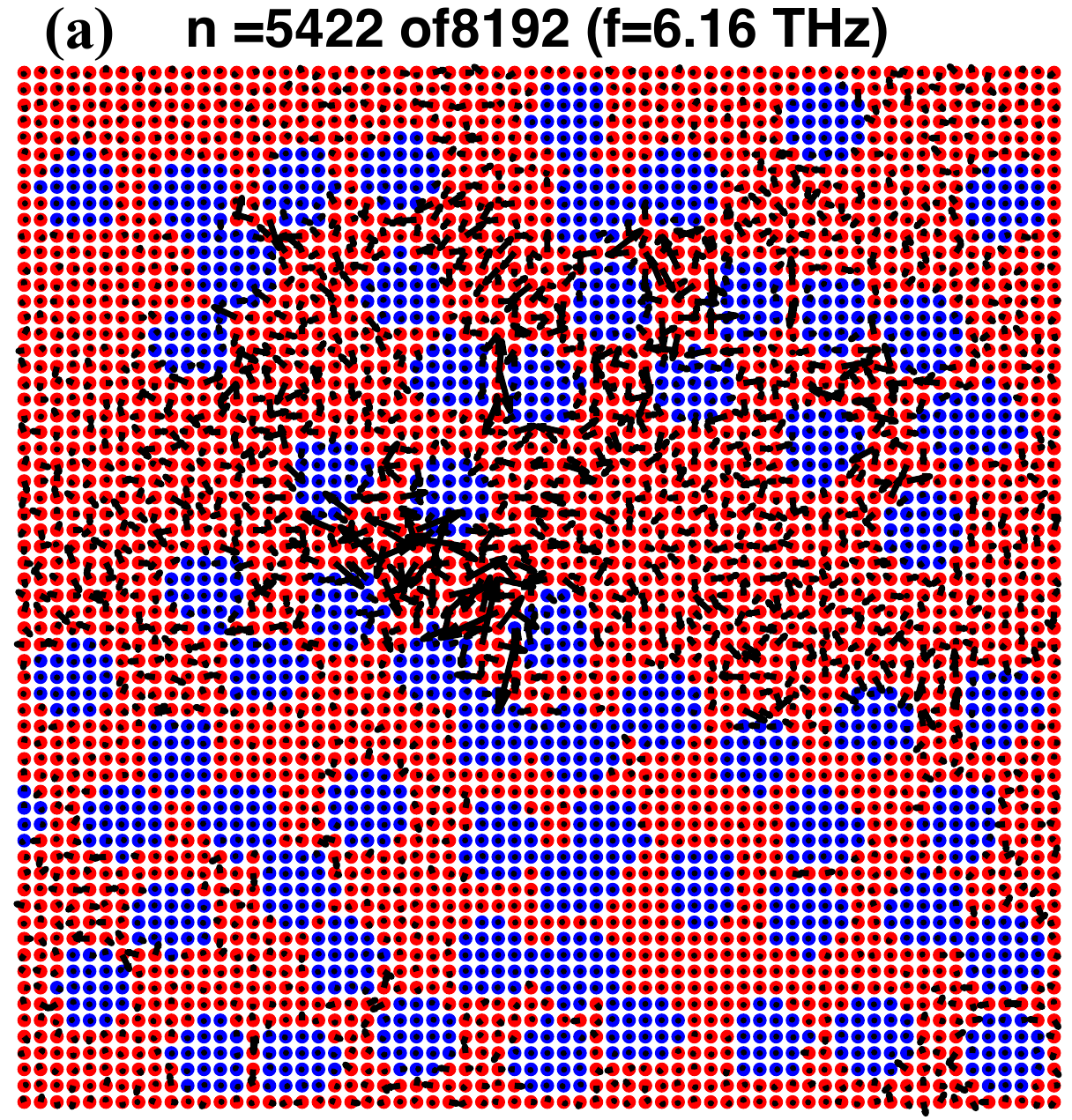}\includegraphics[width=0.4\textwidth]{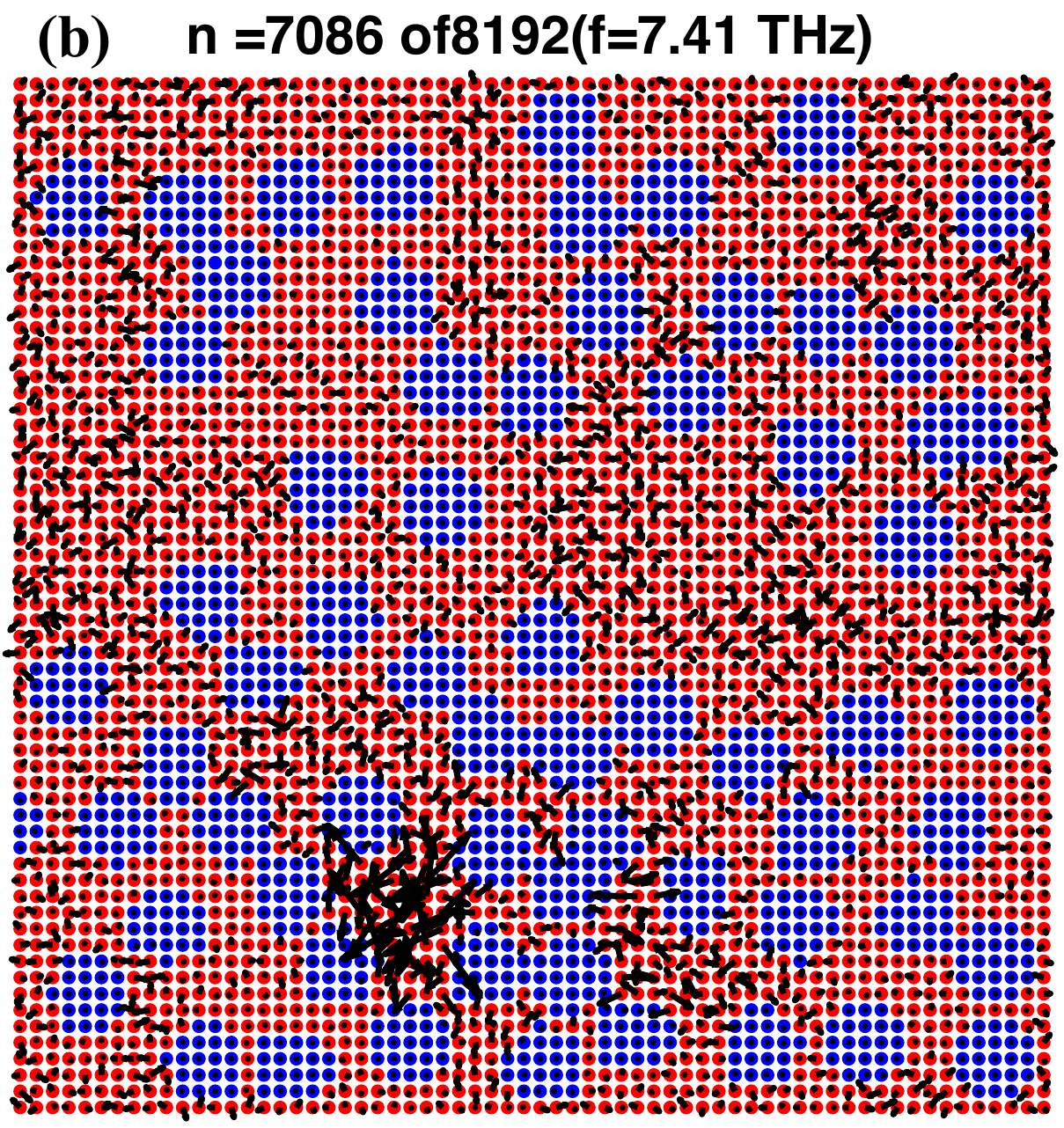}\\
\includegraphics[width=.4\textwidth]{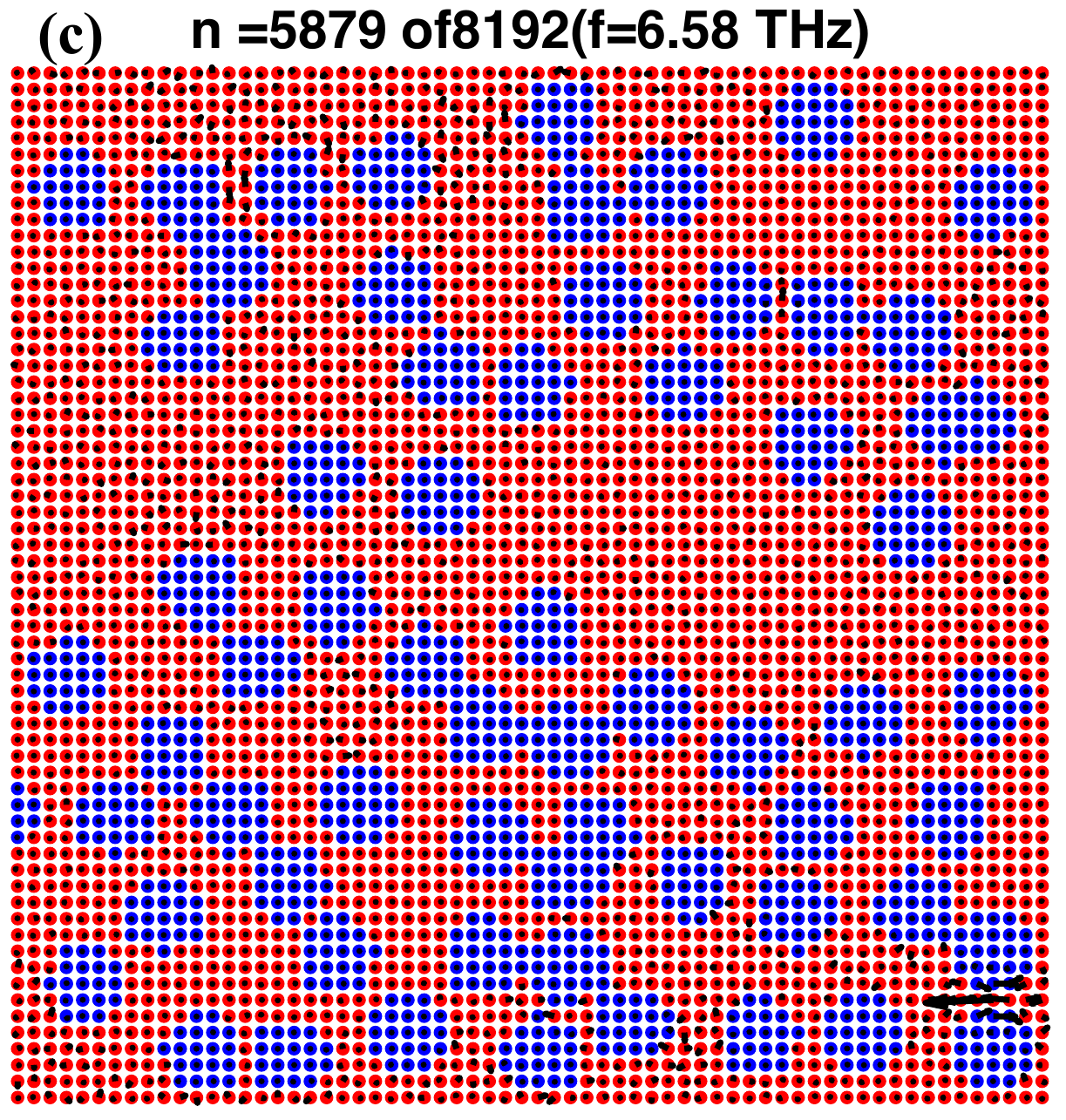}\includegraphics[width=0.4\textwidth]{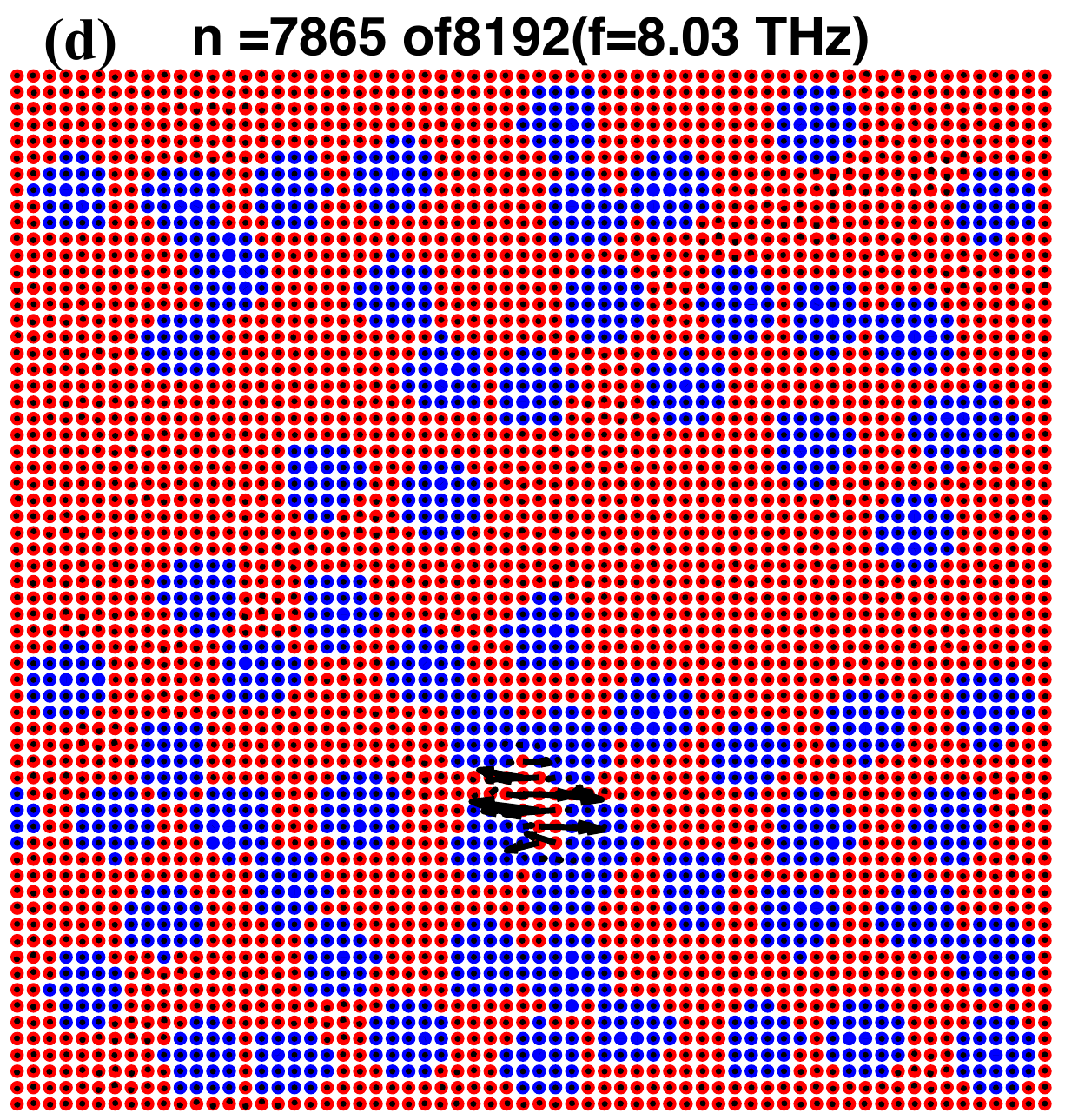}\\
\end{center}
\caption{\label{fig:displacement_patterned_heavyparticle} Example displacement patterns for localized modes.  Additional examples and their fits for localization length are given in the Supporting Materials\cite{supplemental_info}.}
\end{figure*}

Ref. \citenum{doi:10.1080/13642819908223054} suggests determining the localization length for each localized mode by fitting to an exponential decay decay as  $|\epsilon|^2\propto \exp(-2r/\xi)$.  Thus, the localization length can be obtained by taking the eigenvector component with largest amplitude a reference point, ${x_0}$, and plotting $\ln|\epsilon|^2$ vs. distance from ${x_0}$ for each degree of freedom, followed by linear regression to determine localization length.  We have done this, and the results are shown in Fig.~\ref{fig:localization_length_modal_decomp}.   Fig.~\ref{fig:localization_length_modal_decomp}a shows the average localization length as a function of frequency at various volume fractions.  The average is performed by calculating $\xi_m$ for each mode and taking the average of modes within a frequency bin (100 bins are used spaced from 0 to 8.6 THz; if no  localized modes are present in a frequency bin, then no value is shown in Fig.~\ref{fig:localization_length_modal_decomp}a).   As expected, localization length generally decreases as volume fraction is increased; in most cased the localization length is larger than the size of the nanoparticles ($D = 5a =$ 1.37nm).  

However we should caution that, we have found that the exponential form does not fit our localized mode displacement patterns well in almost all cases; still, the approach provides a measure of the spatial extent of the localized modes.  Some example fits are given in the Supplemental Materials\cite{supplemental_info}, but the reason an exponential decay does not fit the data becomes clear upon inspection of some representative localized displacement patterns, shown in Fig.~\ref{fig:displacement_patterned_heavyparticle}.  The “localized” modes in nanoparticle-in-matrix materials are in fact generally found to be somewhat delocalized with an extent defined by a confinement region (i.e. a light mass region nearly encompassed by heavy particles).  In some cases these “localized” regions are large (Fig.~\ref{fig:displacement_patterned_heavyparticle}a\&b) and in other cases they may be constituted of just one or a few atoms trapped between adjacent nanoparticles (Fig.~\ref{fig:displacement_patterned_heavyparticle}c\&d).  In either case, the cause seems not to be an interference effect as in Anderson localization, but instead confinement by a boundary that cannot support the high vibrational frequencies available in the light material.

%

\begin{figure*}
\begin{center}
\includegraphics[width=.80\textwidth]{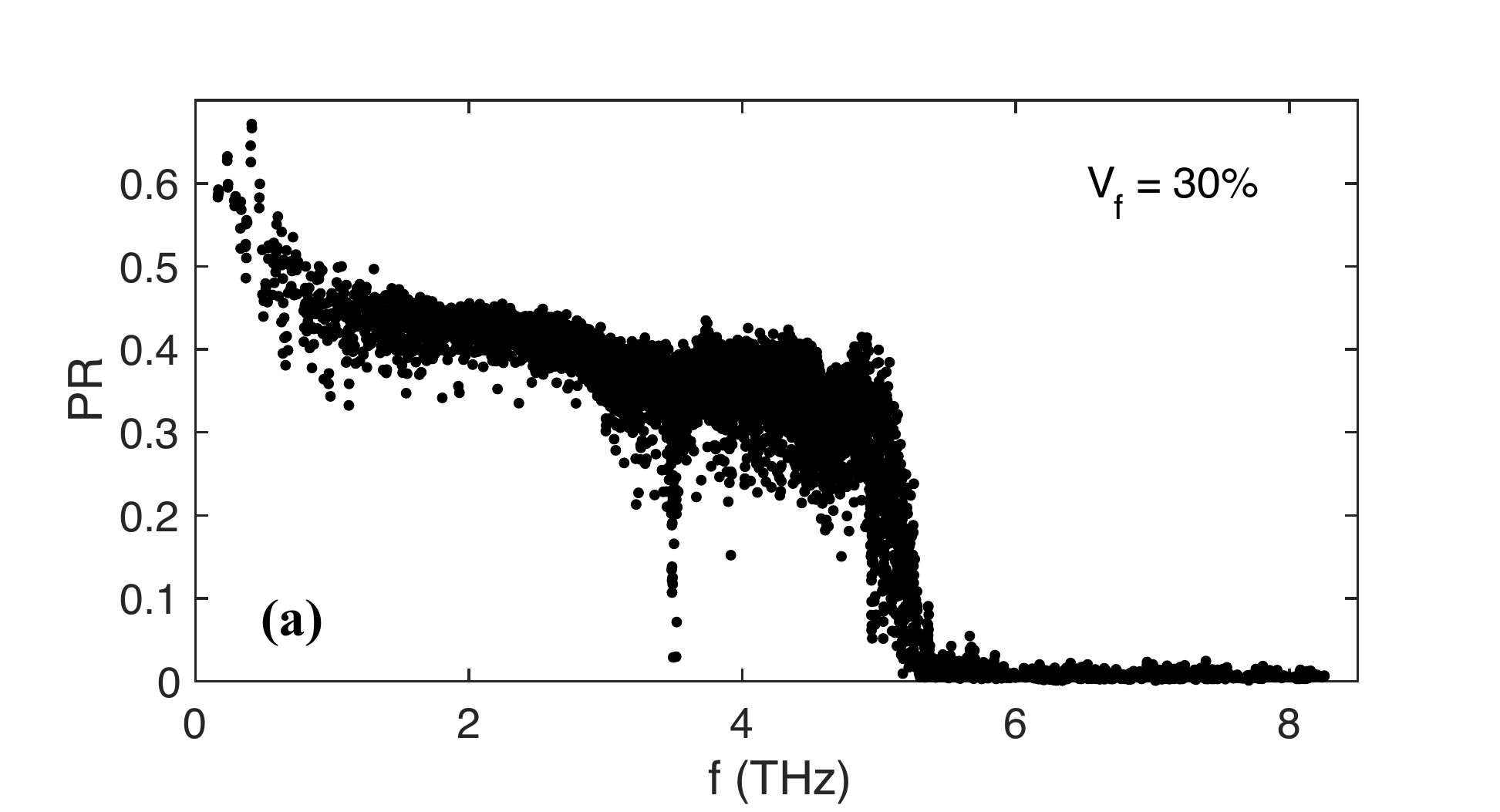}\\
\includegraphics[width=0.40\textwidth]{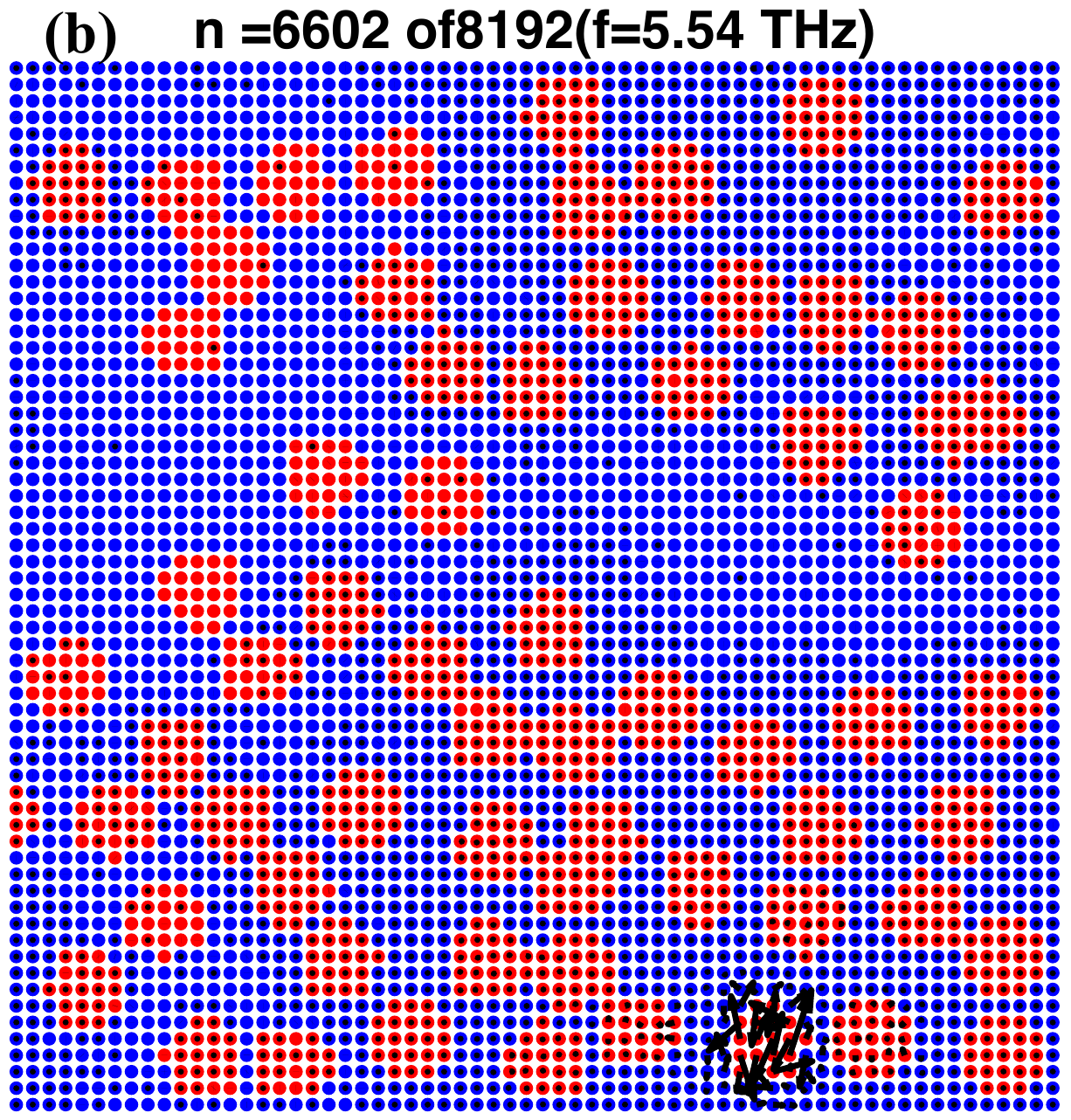}\includegraphics[width=0.40\textwidth]{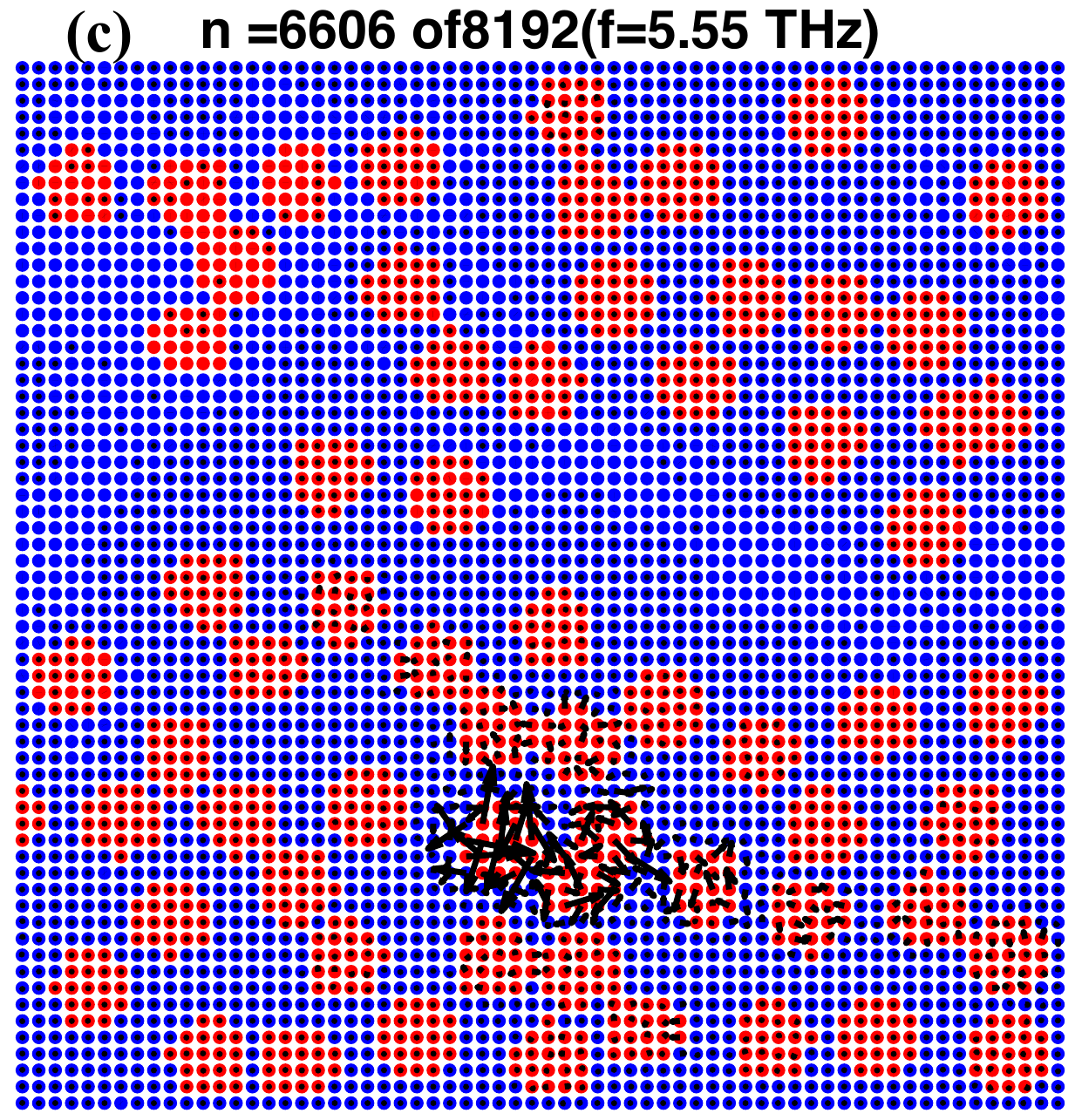}\\
\end{center}
\caption{\label{fig:displacement_patterned_lightparticle} (a) Participation ratio for a 30\% $V_f$ region with heavy matrix and light nanoparticle ($M_1 = 72 m_p$, $M_2 = 28 m_p$); $D = 5a$.  (b \& c) Representative displacement patterns for particular localized modes corresponding to (a).}
\end{figure*}

Based on that rationale, one would expect to find greater localization of modes in a nanoparticle-in-matrix material where the nanoparticles were composed of the lighter material, since then all modes above the highest frequency of the heavier material would be confined within each particle and evanescently decay outside the boundary of the particle.  We explore this hypothesis in Fig.~\ref{fig:displacement_patterned_lightparticle}a, where we perform modal decomposition after reversing the mass of the matrix and particle (i.e. $M_1$ = 72 ${m_p}$, $M_2$ = 28 $m_p$).  It’s observed that $PR$ values for all modes above the maximum frequency of the heavier matrix (cutoff 5.3 THz, see Fig.~\ref{fig:DOSandPR}a) are small.  Examples of the associated displacement patterns are shown in Fig.~\ref{fig:displacement_patterned_lightparticle}b\&c, demonstrating the modes are contained regions confined the light nanoparticles; for low number density regions, the localized modes are confined to individual nanoparticles (\ref{fig:displacement_patterned_lightparticle}b), but regions can also be seen where localized modes spread over several contacting nanoparticles (\ref{fig:displacement_patterned_lightparticle}c).  This implies coupling between modes in adjacent particles is governed by tunneling, which would exponentially increase with distance between particles.  Evidence of this can be seen in Fig. 5 which shows a very rapid increase in localization length as number density is decreased from 48 to 34\% volume fraction.   Unfortunately, a deficiency of the FDPML method, as well as the related NEGF method, is that they cannot easily be used to study transmission coefficient or transmission function through systems where the matrix is heavier than the nanoparticles because the contact region cannot actually support incident modes with such high frequencies (such as those above 5.3 THz in Fig \ref{fig:displacement_patterned_lightparticle}a); an alternative for these methods would be to make the PML/contacting region out of the lighter material, but this introduces other problems as then there is a impedance mismatch and associated reflection associated with the contact itself.  

\section{Conclusion}
In summary, the inverse and exponential scaling of transmission with length in the diffusive and localized regimes, respectively, were used to calculate the mean free path and localization length of phonons in 2D randomly embedded nanoparticle domains. We have demonstrated that mean free path near the Mie regime follows the well-known individual scattering relation, $\ell_{mfp}=1/(\sigma_s \eta)$ relatively well for  volume fractions as large as $\approx30\%$. However, at larger volume fractions, due to multiple scattering effects, the mean free path deviates from this relationship and increases, a phenomenon that occurs sooner for modes with smaller scattering parameter near the Rayleigh regime. By comparing the multiparticle scattering simulations here to predictions made using the one particle scattering cross section, we find that one particle scattering cross section typically underpredicts the observed mean free path by a constant numerical factor.  By computing scattering phase functions in these cases using a modification to the FDPML approach, we attribute the underprediction to strong forward scattering of some modes in the Mie regime. 

We also investigated transport by localized modes using the Landauer method and find that transport by localized modes is only seen at very high frequencies, exceeding the maximum frequencies available by either the heavy particles in the case where heavy particles are embedded in a light matrix or exceeding those of the heavy matrix in the case where light particles are embedded in a heavy matrix.  In the case of heavy embedded nanoparticles, the required volume fractions to observe localized behavior are high with localization lengths that steeply decrease with volume fraction.  Using modal decomposition, we show that the displacement patterns associated with localized modes correspond to regions of light material surrounded or nearly surrounded by heavy materials, which suggests that confinement and exponentially decaying evanescent modes at the edge of the confinement regions govern these modes, rather than Anderson localization.

\begin{acknowledgments}
This material is based upon work supported by the National Science Foundation under Grant No. 1653270.   Computational resources for research was supported in part through the use of the Farber and Caviness computer clusters and associated Information Technologies (IT) resources at the University of Delaware.
\end{acknowledgments}

%

\end{document}